\newcommand{\m}{\rm m}          
\newcommand{\q}{Q}              
\newcommand{\real}{R\hspace{-1.4ex}R}      

\documentstyle[12pt]{iopart}      

\begin{document}

\baselineskip 14pt

\jl{1}

\title[Long-term memory from extremal driving]
{Extremal driving as a mechanism for generating long-term memory}

\author{David Head\footnote[2]{Electronic address: david@ph.ed.ac.uk}}

\address{Department of Physics and Astronomy, University of Edinburgh,
JCMB King's Buildings, Mayfield Road, Edinburgh EH9 3JZ, UNITED KINGDOM.}

\begin{abstract}
    It is argued that systems whose elements are renewed according to an 
    extremal criterion can generally be expected to exhibit long-term memory.
    This is verified for the minimal extremally driven model,
    which is first defined and then
    solved for all system sizes $N\geq2$ and times $t\geq0$,
    yielding exact expressions for the persistence $R(t)=[1+t/(N-1)]^{-1}$
    and the two-time correlation function
    $C(t_{\rm w}+t,t_{\rm w})=(1-1/N)(N+t_{\rm w})/(N+t_{\rm w}+t-1)$.
    The existence of long-term memory is inferred from the scaling of
    $C(t_{\rm w}+t,t_{\rm w})\sim f(t/t_{\rm w})$, denoting {\em aging}.
    Finally, we suggest ways of investigating the robustness of this
    mechanism when competing processes are present.
\end{abstract}

\pacs{05.40.-a, 05.65.+b, 64.70.Pf}

\maketitle

In recent years there has considerable progress in identifying the mechanisms
responsible for long-term memory in glasses and other slowly relaxing systems,
with processes such as domain coarsening and diffusion over random free energy
landscapes now well established
(see {\em e.g.}~\cite{Bouchaud1998,Bouchaud2000,Cugliandolo1993}
and references therein).
However, long-term memory has also been observed in a class of systems,
namely the so-called {\em extremally driven} models, for which there is
no obvious underlying mechanism.
This is not a satisfactory state of affairs, as these models have applications
covering a broad range of physically important situations, in particular to
athermal and low temperature systems such granular media, flux creep {\em etc.}
One may reasonably ask how we can ever expect to understand the real
systems if even their
simplified models behave in a way that cannot be properly explained.

The defining characteristic of extremally driven models is that they
are updated by identifying an `active' region of the system,
and {\em renewing} this region whilst leaving the remainder unaltered.
The active subsystem is chosen according to some kind of {\em extremal} criterion;
often it will be centred on
the location of the minimum (or, equivalently, the maximum)
of some spatially varying scalar variable,
but other possibilities have been considered.
Models that belong to this class include invasion
percolation~\cite{Wilkinson1983},
the Bak-Sneppen model~\cite{Bak1993} and the granular shear model of
T\"or\"ok {\em et al.}~\cite{Torok2000}, amongst others~\cite{Paczuski1996}.
Recently, both the Bak-Sneppen and granular shear models
have been found to exhibit {\em aging}~\cite{Torok2000,Boettcher1997a,Boettcher1997b},
which indicates the existence of some form of long-term memory.
The Bak-Sneppen model, along with many other extremally driven models,
is {\em critical} in the sense that it has been placed by construction
at a critical point ({\em i.e.} a continuous phase transition)
of a broader phase diagram.
By contrast, the granular shear model of T\"or\"ok {\em et al.} is {\em not} critical,
so if the same mechanism is responsible for both cases of aging,
it cannot be due to any of the properties of the critical state.

In this Letter we demonstrate that extremal driving {\em by itself} is enough
generate long-term memory, and claim that this is the true cause
of the aging observed in the Bak-Sneppen and granular shear models.
We further speculate that this mechanism is somewhat {\em robust} and that
many other extremally driven models will also age; to the best of our
knowledge, such behaviour has never been looked for in the other models
in this class.
The central part of this work is the solution of the simplest
extremally driven model, which is shown to age in a similar manner
to that observed in ~\cite{Torok2000,Boettcher1997a}.
Since the only mechanism in the model is the extremal driving, it is
reasonable to infer that this is the cause of the long-term memory.
A secondary aspect of this work is that the model can be
{\em exactly} solved for all system sizes and times.
This allows the finite size effects and transient behaviour to be
explicitly calculated, which is rarely possible in systems exhibiting
slow relaxation.
It is also hoped that this work will help to extend the relationship
between extremal statistics and glassy relaxation
that was originally stressed by Bouchaud 
and M\'ezard~\cite{Bouchaud1997}.

The model to be studied here is defined as follows.
The system consists of $N$ elements which are each assigned a single
scalar variable $x_{i}$\,, $i=1\ldots N$,
drawn from the fixed probability distribution function $p(x)$.
For every time step $t\rightarrow t+1$,
the element with the smallest $x_{i}$ in the system is selected
and {\em renewed} by assigning it a new value of~$x_{i}$\,,
which is drawn from $p(x)$ as before.
Non-degeneracy is assumed, {\em i.e.} no two $x_{i}$ can take the
same value, which is valid as long as $N$ is finite and
$p(x)$ contains no delta-function peaks.
Selecting the maximum rather than the minimum would result in
entirely equivalent behaviour and we henceforth restrict our
attention to the minimum case only.
This minimal model can be viewed as a one-dimensional version
of the granular shear model~\cite{Torok2000},
or equally as the Bak-Sneppen model without interactions~\cite{Bak1993}.

Before solving the model, it is instructive to
describe the emergence of long-term memory in qualitative terms.
For this system, as for any system subjected to extremal driving,
the typical values of $x_{i}$ increase monotonically in time.
This means that any recently renewed element is likely to have a
smaller $x_{i}$ than the bulk, and hence a shorter than average lifespan
until it is again renewed.
Correspondingly, elements that have not been renewed for some time
will have a longer than average life expectancy.
This separation between the shortest and the longest lived elements
will become more pronounced as the system evolves and the average
$x_{i}$ in the bulk increases.
Thus one might reasonably expect a broad distribution of relaxation times,
and the possibility of long-term memory.

To put this picture into a more precise framework,
let $P_{t}(S)$ be the probability to find the system in a state $S$
after $t$ updates, where $S=\{x_{1},x_{2}\ldots x_{N}\}$
[more formally the probability is $P_{t}(S)\prod_{i=1}^{N}{\rm d}x_{i}$
to simultaneously find the first variable in the range
$(x_{1},x_{1}+{\rm d}x_{1})$,
the second in the range $(x_{2},x_{2}+{\rm d}x_{2})$, {\em etc.}]
Only one of the $x_{i}$ changes value during a single time step.
Thus to find $P_{t+1}(S)$ from $P_{t}(S)$, one must integrate
over the region of phase space in which each of the $x_{i}$ is the smallest
and replace it with a value drawn from $p(x)$, {\em i.e.}

\begin{equation}
    P_{t+1}(S)=\sum_{i=1}^{N}p(x_{i})\int_{-\infty}^{\m_{i}}
    P_{t}(S^{(i)})\,{\rm d}x^{\prime}_{i}
\label{e:master}
\end{equation}

\noindent{}where the $\real^{N-1}\rightarrow\real$ function $m_{i}$ is defined by

\begin{equation}
    m_{i} = \min_{j\neq i} \{x_{j}\}
\label{e:defnmi}
\end{equation}

\noindent{}and $S^{(i)}$ is shorthand for
$\{x_{1},\ldots,x_{i-1},x^{\prime}_{i},x_{i+1},\ldots,x_{N}\}$.
The factor of $p(x_{i})$ on the right hand side of (\ref{e:master})
can be removed by making the change of
variables \mbox{$u_{i} = \int_{-\infty}^{x_{i}}p(z){\rm d}z$},
\mbox{$0\leq u_{i}\leq1$}, giving

\begin{equation}
    \q_{t+1}(S) = \sum_{i=1}^{N}\,\int_{0}^{\m_{i}}\,
    \q_{t}(S^{(i)})\,{\rm d}u'_{i}
\label{e:rescaledmaster}
\end{equation}

\noindent{}where $\q_{t}\prod_{i=1}^{N}{\rm d}u_{i}=P_{t}\prod_{i=1}^{N}{\rm 
d}x_{i}$\,.
$S$, $S^{(i)}$ and $m_{i}$ are here defined exactly as in (\ref{e:master})
and (\ref{e:defnmi}), except with the $x_{i}$ replaced by $u_{i}$\,.
Scaling $p(x)$ out of the master equation in this manner reflects that,
as in any extremally driven model, the dynamics
depends only on the {\em order} of the $x_{i}$
and {\em not} their relative spacings
(note that there is no need to invoke universality to prove this result).

Before proceeding to solve the master equation~(\ref{e:rescaledmaster}),
it is useful to state and prove the following identities.
Firstly,

\begin{equation}
    \int_{0}^{m_{i}}m_{j}^{t}\,{\rm d}u_{i}=\left\{
    \begin{array}{l@{\quad:\quad}l}
	m_{i}^{t+1} & i=j\quad, \vspace{0.3cm}\\
	\displaystyle{\frac{m_{i}^{t+1}}{t+1}} & i\neq j\quad.
    \end{array}
    \right.
    \label{e:id1}
\end{equation}

\noindent{}The $i=j$ case is trivial (since $m_{j}$ is independent
of $u_{j}$), whereas for $i\neq j$ observe that $m_{j}\equiv u_{i}$ over
the entire range of integration, from which the result follows.
Another useful identity is

\begin{equation}
    \int_{{\mathcal{D}}_{i}}m_{j}^{t}\,{\rm d}V = \left\{
    \begin{array}{c@{\quad:\quad}l}
	\displaystyle{\frac{(N-1)!(t+1)!}{(N+t)!}} & i=j\quad, \vspace{0.3cm}\\
	\displaystyle{\frac{(N-1)!\,t!}{(N+t)!}} & i\neq j\quad,
    \end{array}
    \right.
    \label{e:id2}
\end{equation}

\noindent{}where ${\mathcal{D}}_{i}$ is the domain of space in which
$u_{i}$ is the smallest, and ${\rm d}V=\prod_{k=1}^{N}{\rm d}u_{k}$\,.
This can be proven by considering in turn each of the $(N-1)!$ subregions
of ${\mathcal{D}}_{i}$ defined by 
$u_{i}<u_{l_{1}}<u_{l_{2}}<\ldots<u_{l_{N-1}}$,
where $l_{k}\neq i$ $\forall k$.
For each permutation of the $l_{k}$\,, the integral limits for each of the
${\rm d}u$ can be inserted and the integration evaluated.
The final result (\ref{e:id2}) then follows from summing over all
the possible permutations.

The rescaled master equation (\ref{e:rescaledmaster}) can be solved 
inductively from the initial state $\q_{0}=1$ by using the first
identity (\ref{e:id1}), giving

\begin{equation}
    \q_{t}(S) = \frac{(N+t-1)!}{t!\,N!}\,\sum_{i=1}^{N}\,\m_{i}^{t}\quad.
    \label{e:gensoln}
\end{equation}

\noindent{}That this is correctly normalised can be checked using
the second identity~(\ref{e:id2}).
Since $\q_{t}$ is symmetric in the $m_{i}$ and therefore the $u_{i}$\,,
the probability that any particular element in the system, say $u_{k}$\,,
is the smallest at a given time $t_{\rm w}$ is just~$1/N$.
However, suppose it is known that $u_{k}$ is {\em not} the smallest at~$t_{\rm w}$\,.
Then $\q_{t_{\rm w}+1}$ can then be found by setting $\q_{t_{\rm w}}$ to zero
in ${\mathcal{D}}_{k}$\,, renormalising, and iterating
(\ref{e:rescaledmaster}) once.
This three-step procedure can be repeated to find the following expression for
$\q^{k}_{t_{\rm w}+t,t_{\rm w}}(S)$, defined as the probability to find the
system in a state $S$ at a time $t_{\rm w}+t$ given that $u_{k}$ was not the minimum at any
of the times $t_{\rm w},t_{\rm w}+1,\ldots,t_{\rm w}+t-1$,

\begin{equation}
    \q^{k}_{t_{\rm w}+t,t_{\rm w}}(S) = \frac{1}{N-1}
    \left(
    \begin{array}{c}
	N+t_{\rm w}+t-1 \\
	N-1
    \end{array}
    \right)
    \sum_{i\neq k}m_{i}^{t_{\rm w}+t} \quad,\quad t\geq1\:.
    \label{e:qreturn}
\end{equation}

\noindent{}The corresponding probability that $u_{k}$ is the smallest,
denoted here by $q^{k}_{t_{\rm w}+t,t_{\rm w}}$, can be calculated by integrating
(\ref{e:qreturn}) over ${\mathcal{D}}_{k}$ and using (\ref{e:id2}),

\begin{equation}
    q^{k}_{t_{\rm w}+t,t_{\rm w}}=\frac{1}{N+t_{\rm w}+t}
    \label{e:preturn}
\end{equation}

\noindent{}which is independent of $k$.
This demonstrates that the probability of an element being renewed
decreases with the time since it was {\em last} renewed,
according to $q^{k}_{t_{\rm w}+t,t_{\rm w}}\sim t^{-1}$\,.
Note that $q^{k}_{t_{\rm w}+t,t_{\rm w}}$ also decreases with $t_{\rm w}$\,.

We are now in a position to calculate the physical quantities
of interest, starting with the persistence $R(t)$~\cite{Derrida1994,Bray1994}.
$R(t)$ is defined as the probability that a randomly chosen
element $i$ has the same value of $x_{i}$ at time $t$ that it had at $t=0$.
Clearly, $R(0)=1$ and $R(1)=(N-1)/N$.
For $t\geq2$, observe that
$R(t)$ only decreases when an element is renewed for the first time,
so $R(t+1)=(1-q^{k}_{t,0})\,R(t)$ and hence from 
(\ref{e:preturn}),

\begin{eqnarray}
    R(t)&=&R(1)\prod_{s=1}^{t-1}\left( 1-q^{k}_{s,0} \right) \\
    &=&\frac{N-1}{N+t-1} \label{e:exactR}\\
    &\sim&\left(\frac{t}{N-1}\right)^{-\theta}
    \label{e:persistence}
\end{eqnarray}

\noindent{}which defines the persistence exponent~$\theta=1$.
The slow decay of $R(t)$ shows that a significant proportion
of the system will remain in its initial state until arbitrarily
late times, already suggesting some form of long-term memory.
Note that there is no cut-off for finite system sizes;
$R(t)$ asymptotically decays algebraically even for $N=2$,
as long as one averages over all possible initial conditions and histories.

The existence of aging can be most clearly expressed in terms of the
two-time correlation function
$C(t_{\rm w}+t,t_{\rm w})$ between the state
of the system at times $t_{\rm w}$ and $t_{\rm w}+t$.
A suitable $C(t_{\rm w}+t,t_{\rm w})$ for this model is the probability that a
randomly chosen element has the same value of $x_{i}$
at $t_{\rm w}$ and $t_{\rm w}+t$
[so $C(t,0)\equiv R(t)$].
As before, $C(t_{\rm w},t_{\rm w})=1$, $C(t_{\rm w}+1,t_{\rm w})=(N-1)/N$ and

\begin{eqnarray}
    C(t_{\rm w}+t,t_{\rm w})&=&
    C(t_{\rm w}+1,t_{\rm w})\prod_{s=t_{\rm w}+1}^{t_{\rm w}+t-1}
    \left( 1-q^{k}_{s,t_{\rm w}} \right)\\
    \label{e:exactC}
    &=& \displaystyle{\frac{N-1}{N}
    \left(\frac{N+t_{\rm w}}{N+t_{\rm w}+t-1}\right)}
    \quad,\hspace{4ex}t\geq1.
\end{eqnarray}

\noindent{}After a short transient this scales as

\begin{equation}
    C(t_{\rm w}+t,t_{\rm w})
    \approx\frac{N-1}{N}\left(
    1+\frac{t}{t_{\rm w}}
    \right)^{-1}
    \quad,\hspace{4ex}\frac{t_{\rm w}}{N}\gg 1.
    \label{e:scaling}
\end{equation}

\noindent{}That $t$ and $t_{\rm w}$ only appear in the ratio $t/t_{\rm w}$
is what we mean by {\em aging}.
Finally, note that in the limit $N\rightarrow\infty$,
the $N$-dependence of (\ref{e:exactR}) and (\ref{e:exactC})
can be removed by renormalising the timescale to
$\tau\equiv t/N$, giving $R(\tau)=(1+\tau)^{-1}$ and
$C(\tau_{\rm w}+\tau,\tau_{\rm w})=(\tau_{\rm w}+1)/(\tau_{\rm w}+\tau+1)$,
respectively.

We have now achieved what we set out to do,
{\em i.e.} demonstrate that even the simplest extremally driven model
has long-term memory, as evident from the aging of
$C(t_{\rm w}+t,t_{\rm w})$ in~(\ref{e:scaling}),
and the slow decay of $R(t)$~(\ref{e:exactR}).
From this we infer that the extremally driven renewal is responsible
for the aging observed in~\cite{Torok2000,Boettcher1997a,Boettcher1997b}
and speculate that other extremal models, such as invasion percolation,
may also age in a similar fashion.
However, it should be stressed that not all extremally driven models will
necessarily exhibit aging.
Indeed, it is already known that including noise-like terms,
by renewing randomly selected elements as well as the extremal one,
introduces an exponential cut-off to the relaxation times and
destroys the long-term memory
(this situation is realised in the mean-field version of the
Bak-Sneppen model, for instance~\cite{Head2000}).
Thus a useful step forward from this work might be to consider modified
versions of the model, to see what physical processes
may enhance or disrupt the effects of extremal driving.
To this end, we tentatively suggest that the following modifications might be
particularly worthy of investigation:
introducing quenched disorder [by assigning each element its own
generating distribution $p_{i}(x_{i})$],
allowing the values of $x_{i}$ before and after renewal to be correlated,
and letting the time step between updates to depend on the $x_{i}$\,.
It is especially hoped that these and other enhancements could be treated
within the exact framework developed here.

In summary, we have argued that systems which are renewed according
to an extremal criterion should be expected to exhibit long-term memory,
and have supported this claim
by showing that even the minimal extremally driven model ages.
Expressions were found for the persistence $R(t)$ and the
two-time correlation function $C(t_{\rm w}+t,t_{\rm w})$ which
corroborate these claims.
Finally, we note that this work also constitutes an instance where the
extremal properties of a system of correlated random variables can
be exactly computed.
To our knowledge, this makes it one of few such systems known
\cite{Majumdar2000preprint,Gumbel1958,Callisto1988}.

\section*{Acknowledgements}
This work was funded under UK EPSRC grant no. GR/M09674.



\Bibliography{99}

\bibitem{Bouchaud1998} Bouchaud J-P, Cugliandolo L F , Kurchan J
and M\'ezard M 1998 {\em Spin glasses and random fields}
ed Young A P (Singapore: World Scientific)

\bibitem{Bouchaud2000} Bouchaud J-P 2000
{\em Soft and fragile matter: Non-equilibrium dynamics, metastability and
flow} ed Cates M E and Evans M R (Bristol: IOP Publishing and
Edinburgh: SUSSP Publications)

\bibitem{Cugliandolo1993} Cugliandolo L F and Kurchan J
1993 {\em Phys. Rev. Lett.} {\bf 71} 173

\bibitem{Wilkinson1983} Wilkinson D and Willemsen J F 1983
{\em J. Phys. A} {\bf 16} 3365

\bibitem{Bak1993} Bak P and Sneppen K 1993
{\em Phys. Rev. Lett.} {\bf 71} 4083

\bibitem{Torok2000} T\"or\"ok J, Krishnamurthy S, Kert\'esz J and Roux S 2000
{\em Phys. Rev. Lett.} {\bf 84} 3851

\bibitem{Paczuski1996} Paczuski M, Maslov S and Bak P 1996
{\em Phys. Rev. E} {\bf 53} 414

\bibitem{Boettcher1997a} Boettcher S and Paczuski M 1997
{\em Phys. Rev. Lett.} {\bf 79} 889

\bibitem{Boettcher1997b} Boettcher S 1997
{\em Phys. Rev. E} {\bf 56} 6466

\bibitem{Bouchaud1997} Bouchaud J-P and M\'ezard M 1997
{\em J. Phys. A} {\bf 30} 7997

\bibitem{Derrida1994} Derrida B, Bray A J and Godr\`eche C 1994
{\em J. Phys. A} {\bf 27} L357

\bibitem{Bray1994} Bray A J, Derrida B and Godr\`eche C 1994
{\em Europhys. Lett.} {\bf 27} 175

\bibitem{Head2000} Head D A 2000
{\em J. Phys. A} {\bf 33} 465

\bibitem{Majumdar2000preprint} Majumdar S N and Krapivsky P L 2000
unpublished ({\em cond-mat/0006236})

\bibitem{Gumbel1958} Gumbel E J 1958 {\em Statistics of extremes}
(New York: Columbia University Press)

\bibitem{Callisto1988} Callisto E 1988
{\em Extreme value theory in engineering} (San Diego: Academic Press)


\endbib
\end{document}